\documentclass[10pt, preprint2]{aastex}

\begin{document}

\slugcomment{Ap. J., Submitted}
\shortauthors{Mori, Walker \&\ Ohishi}
\shorttitle{Dense clouds and UID EGRET sources}

\title{Dense gas clouds and the Unidentified EGRET sources}

\author{Mark Walker$^{1,2}$, Masaki Mori$^3$ and Michiko Ohishi$^3$}

\affil{1. School of Physics, University of Sydney, NSW 2006, Australia}
\affil{2. Australia Telescope National Facility, CSIRO}
\affil{3. Institute of Cosmic-Ray Research, University of Tokyo, Kashiwa, Chiba 277-8582, Japan}

\begin{abstract}
Cold, dense gas clouds have been proposed as a
major component of the Galactic dark matter; such
clouds can be revealed by the gamma-ray emission
which arises from cosmic-rays interacting with the
gas. If this dark matter component is clustered then
highly luminous GeV sources result, preferentially at
low to mid Galactic latitudes where they lie within
the cosmic-ray disk. The predicted emission for such
clusters is steady, continuum emission with peak power
emerging at several hundred~MeV.  These sources would
not have obvious counterparts at other wavelengths and
are thus of interest in connection with the UnIDentified
(UID) EGRET sources. Here we present a Monte Carlo simulation
of the gamma-ray source population due to cold gas clouds,
assuming a Cold-Dark-Matter-like mass spectrum for the
clustering. We find that $\sim280$ EGRET sources are
predicted by this model, with a median Galactic latitude
of $12^\circ$ for the population, and a median angular size
of $2.2^\circ$. The latitude and size distributions are
consistent with the UID EGRET source data, but the
source counts are clearly overpredicted. On the basis
of these results we propose that clusters of cold gas
clouds comprise the majority population of the observed
UID EGRET sources. Our interpretation implies that there
should be microwave counterparts to most of the UID sources,
and will thus be strongly constrained by data from the
present generation of microwave anisotropy experiments.
\end{abstract}

\keywords{galaxies: Milky Way --- galaxies: halos
--- dark matter --- gamma-rays: general --- cosmic-rays: general
--- ISM: clouds}

\section{Introduction}
The nature of the UnIDentified (UID) EGRET sources
(Hartman et al 1999) is one of the most interesting
puzzles in contemporary high energy astrophysics. This
puzzle persists primarily because of the dimensions
-- typically of order a degree -- of the gamma-ray error
boxes: large uncertainties in source position prohibit
deep searches for counterparts at other wavelengths.
We know, however, that this is only part of the
problem: in some cases the EGRET source is bright
enough that it can be located quite accurately, yet still
no clear counterpart can be found (e.g. Mirabel et al 2000;
Reimer et al 2001).

The majority of the discrete sources detected by
the EGRET instrument  are UID sources (Hartman et al
1999); it is not clear at present whether they are
fundamentally similar to the known GeV emitters, or
whether they represent an entirely new population.
Because the UID sources are predominantly located at
low to mid Galactic latitudes, they must be a Galactic
population. This rules out one of the major classes
of known discrete celestial GeV gamma-ray sources,
namely the flat-spectrum radio quasars as these are
extragalactic and hence isotropically distributed on
the sky. Much attention has therefore focussed on the
possibility that the UID sources are related to massive
stars and their evolutionary end-points. In particular
there has been intense interest in young pulsars
as a contributor to the UID population  (e.g. Bailes
and Kniffen 1992; Kaaret and Cottam 1996; Roberts,
Romani and Johnston 2001; Grenier and Perrot 2001);
several of the identified GeV sources are young
pulsars (Hartmann et al 1999). However, the observed
latitude distribution of the UID sources is too broad
to be readily compatible with young pulsars (Yadigaroglu
and Romani 1997). Supernova remnants (Sturner, Dermer
and Mattox 1996; Torres et al 2003), molecular clouds
and massive star clusters (Montmerle 1979; Benaglia
et al 2001) share this difficulty; but Gehrels et al
(2001) have noted that the sky distribution of UID
sources is consistent with a strong contribution
from objects in Gould's Belt, thus supporting a
connection with massive stars.

Given the difficulty of detecting counterparts to the
UID EGRET population, it is natural to consider the possibility
that these sources may be connected with the dark matter
problem. The currently popular cosmological model, the
Cold Dark Matter model (e.g. Peebles 1993),
stipulates that the dark
matter is composed of weakly interacting particles; it
is possible that these particles are massive -- i.e.
they are WIMPS -- and could yield gamma-rays by
annihilation or decay. This possibility has been explored
by many authors, e.g. Calc\'aneo-Rold\'an and Moore (2000),
Bergstr\"om, Edsj\"o and Gunnarsson (2001).

There are of course other types of dark matter. Of particular
interest in the present context is baryonic dark matter in the
form of cold, dense gas clouds: unlike diffuse gas, cold, dense
clouds are difficult to detect directly (Pfenniger, Combes and
Martinet 1994; Gerhard and Silk 1996; Combes and Pfenniger 1997;
Walker and Wardle 1999). There are many astrophysical motivations
for considering dark matter in this form, ranging from
the properties of late-type galaxies (Pfenniger, Combes
and Martinet 1994; Walker 1999), to observations of radio-wave
scintillation (Henriksen and Widrow 1995; Walker and Wardle 1998),
and the ``Blank Field'' SCUBA sources (Lawrence 2001). 
Indirect constraints -- from the small amplitude of
fluctuations in the Cosmic Microwave Background (CMB), and
from Big Bang Nucleosynthesis  -- are generally supposed
to exclude any significant amount of baryonic dark matter
(e.g. Turner and Tyson 1999), but these arguments are not
entirely free of loopholes (Hogan 1993; Walker and Wardle
1999) and direct constraints are desirable. There is not
yet any direct observational evidence which excludes a
substantial fraction of the Galactic dark matter being in
the form of cold, dense gas clouds (Pfenniger, Combes and
Martinet 1994; Gerhard and Silk 1996; Walker and Wardle
1999; Rafikov and Draine 2000; Ohishi, Mori and Walker
2003; Walker and Lewis 2003). But the key motivation here
for considering this form of dark matter is that gas clouds
are expected to emit gamma-rays if they reside in
the cosmic-ray disk (e.g. Bloemen 1989), and any unexplained
gamma-ray flux might therefore be a signature of dark matter
in this form (De~Paolis et al 1995; Kalberla, Shchekinov and
Dettmar 1999; Sciama 2000). The Galactic gamma-ray halo
reported by Dixon et al (1998) has, for example, been
interpreted in these terms (De~Paolis et al 1999). The
low mass ($\sim10^{-4}\,{\rm M_\odot}$) of the individual
clouds renders them undetectable by EGRET when
they exist in isolation (see \S5).
However,  clustering is ubiquitous in gravitating systems,
and we therefore consider the possibility that the UID EGRET
sources are simply clusters of cold, dense gas clouds. 

The present paper is organised as follows: we begin by
describing the model we have employed for the distribution
of cold gas clouds (\S2), and the emission which is
expected from the gas as a result of cosmic-ray
interactions (\S3); our model cosmic-ray distribution
is given in \S4; we present the results of our Monte Carlo
simulation in \S5, and then compare these results to
the EGRET data in \S6.

\section{Dark matter model}
To proceed with a calculation we need to specify a model
for the clustering and distribution of the dark matter.
The success of the modern structure formation paradigm,
exemplified by the Cold Dark Matter (CDM) model
(e.g. Blumenthal et al 1984; Davis et al 1985; Peebles
1993), argues that any acceptable
model must possess clustering properties that are
similar to those of CDM, in order to match the
data on large-scale-structure. In CDM simulations it is
found that the dark matter within individual galaxy halos
is clustered into mini-halos, with roughly equal contributions
to the total dark matter density being made by all mini-halo
mass scales (Kauffmann, White and Guideroni 1993, Moore et al
1999; Klypin et al 1999) --- i.e.
${\rm d}n/{\rm d}M\propto M^{-2}$, for mini-halos of mass
$M$ and number density $n$. Following Walker, Ohishi and Mori
(2003: WOM03 hereafter), we adopt this mass spectrum
for our calculations, spanning the mini-halo mass range
$0.1\le M/{\rm M_\odot}\le10^{10}$. We further assume that
all of the Galactic dark matter is in the form of cold gas
clouds, clustered into mini-halos. Numerical simulations
of structure formation typically show only $\sim\,$10\% of
the dark matter within a halo to be clustered into mini-halos
(Ghigna et al 1998; Klypin et al 1999). However, these
simulations do not have the resolution required to find
low mass ($M<10^8\;{\rm M_\odot}$) mini-halos, and are
relevant to only the top 20\% (mass fraction) of our adopted
spectrum. Assuming that the true clustering spectrum does
indeed extend to $M\ll10^8\;{\rm M_\odot}$,
we thus expect that the total mass fraction in mini-halos
is actually several times larger than is revealed by
simulations. We have therefore adopted the extreme model
in which all of the dark matter is clustered; this has
the virtue of being a limiting case.

The adopted mini-halo mass spectrum is assumed to be
the same throughout the Galaxy, but the normalisation
(i.e. total number density of mini-halos) varies in
direct proportion to the mean dark matter density of
the Galaxy. Gas clouds are a collisional form of dark
matter, so that an initially singular isothermal halo
develops a core whose radius increases with
time (Walker 1999). Simple calculations of this evolution
lead to a preferred value for the column density of the
individual gas clouds of $\Sigma\simeq140\,{\rm g\,cm^{-2}}$,
and hence a preferred model for the Galactic dark matter
density distribution. We adopt this model, namely an
isothermal sphere, with circular speed of
$220\;{\rm km\,s^{-1}}$, and a core radius of
6.25~kpc (Walker 1999). 

The density profile of the individual mini-halos themselves
is also of some interest to us, in that it determines the
apparent structure of the gamma-ray sources. At present there
is little motivation for a detailed appraisal of these
properties. The key point, however, is that the mini-halos
are not expected to be point-like gamma-ray sources, and
we are therefore interested in estimating their sizes.
We follow WOM03 in assuming (see Ghigna
et al 1998; Moore et al 1999) that the mini-halos are
tidally truncated at perigalacticon, with the location of
the latter estimated as one half of the Galactocentric
radius of the mini-halo. This leads to gamma-ray sources
with estimated angular sizes of order degrees.

\section{Gamma-ray emissivity}
For high column-density gas clouds, predicting the effects
of cosmic-ray bombardment is more complicated than for
diffuse gas because the cosmic-ray spectrum is substantially
modified by transport through the cloud (Kalberla, Shchekinov
and Dettmar 1999; Sciama 2000).  Scattering and absorption
of the emergent gamma-rays is similarly important. We have
therefore used the Monte Carlo event simulator
GEANT\footnote{{\tt http://wwwinfo.cern.ch/asd/geant}} to
estimate the gamma-ray spectrum resulting from cosmic-ray
bombardment of dense gas clouds (Ohishi, Mori and Walker 2003).
By using GEANT we are able to study the transport of
cosmic-rays into dense gas clouds, and their resulting
gamma-ray emission, with all relevant interactions included.
We employ the results of GEANT simulations for clouds
of mean column-density $100\,{\rm g\,cm^{-2}}$, this
being the closest, simulated case to our preferred
value of $\Sigma$ (\S2). For the most part we shall
need only the integrated gamma-ray emissivity of the
gas clouds. In the solar neighbourhood we estimate the
$>100$~MeV emissivity due to cosmic-ray hadrons (mainly
protons) to be $J_p=5.1\times10^{-2}\;{\rm ph\,g^{-1}s^{-1}}$,
while that due to cosmic-ray electrons is
$J_e=5.2\times10^{-3}\;{\rm ph\,g^{-1}s^{-1}}$
(Ohishi, Mori and Walker 2003). Here we have adopted
the median cosmic-ray proton spectrum of Mori (1997),
and the cosmic-ray electron spectrum of Skibo and
Ramaty (1993). The total emissivity is thus estimated to
be $J=J_p+J_e=5.6\times10^{-2}\;{\rm ph\,g^{-1}s^{-1}}$.

\section{Cosmic-rays}
Most of our information on the Galactic cosmic-ray
spectrum comes from direct observation of energetic particles
in the solar neighbourhood, and we know very little
about the spectrum at other locations in the Galaxy.
Although it is possible to construct theoretical models
of the cosmic-ray spectrum and distribution throughout
the Galaxy, based on assumed sources, sinks and diffusive
particle propagation (e.g. Porter and Protheroe 1997,
Strong, Moskalenko and Reimer 2000),
such models introduce an additional layer
of complexity which is unwarranted in the present context.
Here we adopt the simple assumption that the shape of the
cosmic-ray spectrum is the same throughout the Galaxy.

It then remains to specify the cosmic-ray energy-density
as a function of position in the Galaxy. Webber, Lee and
Gupta (1992: WLG92) constructed numerical models of
cosmic-ray propagation in the Galaxy; they did not give
any analytic forms for their model cosmic-ray distributions,
but an appropriate analytic approximation can be
deduced from the results which they obtained. They found
that the cosmic-ray radial distribution reflects,
in large part, the radial dependence of cosmic-ray
sources, with a modest smoothing effect introduced
by diffusion. We have therefore adopted WLG92's preferred
model (their model \#3) for the radial distribution of
sources as our model for the radial distribution of
cosmic-rays.

The various spectra of cosmic-ray isotope ratios
considered by WLG92 favour models in which the diffusion
boundaries are in the range $2-4$~kpc above and below
the plane of the Galaxy. We adopt the midpoint of this range.
WLG92 do not give a simple functional form for the vertical
variation of cosmic-ray density within this  zone, so we have
simply assumed an exponential model: $\exp(-|z|/h)$. We
know that in WLG92's models the cosmic-ray density is fixed
at zero at the diffusion boundaries, and consequently the
scale-height of the exponential should be approximately half of
the distance to the diffusion boundary, i.e. $h\simeq1.5$~kpc.

These considerations lead
us to the model cosmic-ray density distribution $U(R,z)$:
\begin{equation}
{{U}\over{U_\odot}}=\left({R\over{R_0}}\right)^{0.6}
 \exp[(R_o-R)/\varrho -|z|/h],
\end{equation}
in terms of cylindrical coordinates $(R,z)$.  Here
$R_o\simeq8.5$~kpc is the radius of the Solar circle,
while $\varrho=7$~kpc, and $h=1.5$~kpc;
$U_\odot=U(R_o,0)$ is the cosmic-ray density in the
Solar neighbourhood. This distribution has the character
of a disk with a central hole.

\begin{figure}
\plotone{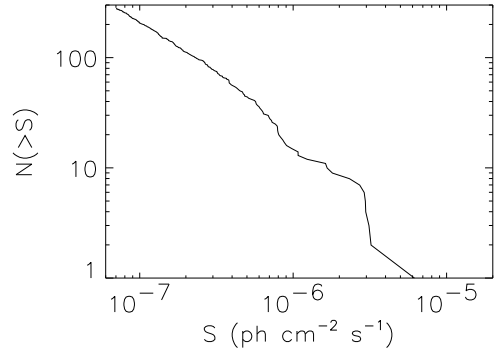}
\caption{The gamma-ray source counts, as a function of flux
(above 100~MeV), for the simulated mini-halo population.}
\end{figure}
\begin{figure}
\plotone{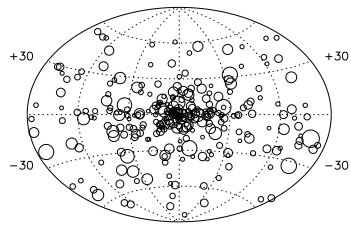}
\caption{Sky distribution of simulated gamma-ray sources
lying above the EGRET flux limit
($7\times10^{-8}\;{\rm ph\,cm^{-2}\,s^{-1}}$ above 100~MeV).
The size of the circle plotted for each source increases
according to the source flux.}
\end{figure}
\begin{figure}
\plotone{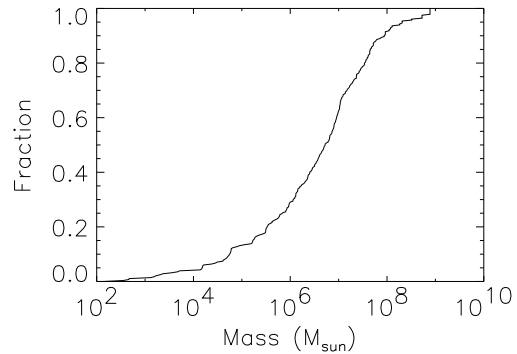}
\caption{The distribution of mini-halo mass, for
the synthetic population of sources which lie above the
EGRET flux limit. The median value is
$5\times10^6\;{\rm M_\odot}$.}
\end{figure}

\section{Results}
Using the model described above, we simulated
$\sim2\times10^8$ halos with masses
$10^2\le M/{\rm M_\odot}\le10^{10}$ located within
50~kpc of the Galactic Centre. The model mini-halo mass
spectrum which we utilised extends to lower masses
($0.1\;{\rm M_\odot}$); these very-mini-halos
(pico-halos?) were not included in the simulation because
a negligible fraction of them can be detected (see later).
By the same token, the individual gas clouds themselves
($M\sim10^{-4}\;{\rm M_\odot}$) are not expected to
be detectable by EGRET. In particular, if all of the
dark matter is assumed to be in a spherical halo of
{\it unclustered\/} clouds, then the closest example
to the Sun should lie at a distance of order 0.1~pc,
and have a flux of order
$7\times10^{-9}\,{\rm ph\,cm^{-2}\,s^{-1}}$ above 100~MeV;
this is roughly an order of magnitude below the detection
limit of EGRET.

\begin{figure}
\plotone{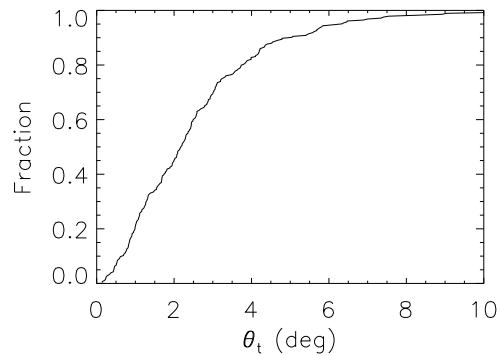}
\caption{The distribution of source radii (tidal radii), for
the synthetic population of sources which lie above the
EGRET flux limit. The median value is 2.2~degrees.}
\end{figure}

\begin{figure}
\plotone{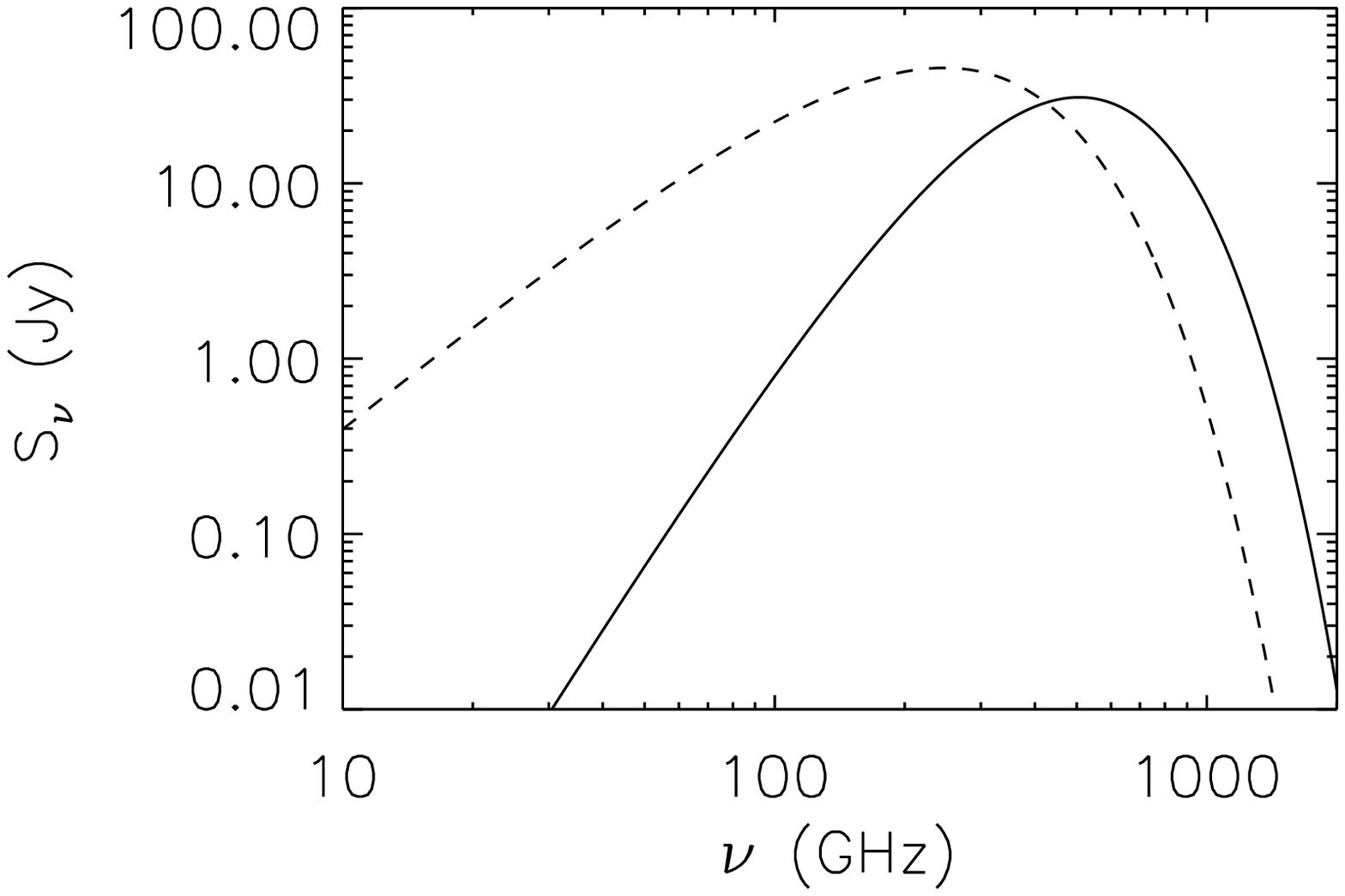}
\caption{Model mini-halo thermal spectra.
The dashed line shows a blackbody spectrum at $T=4.2$~K,
and the solid line shows a ``dusty'' ($\alpha=2$) spectrum
at $T=4.9$~K. The thermal flux ($S_\nu$)
scales in direct proportion to the gamma-ray
flux, and this plot is made for a gamma-ray flux
of $10^{-7}\;{\rm ph\,cm^{-2}s^{-1}}$ above 100~MeV.}
\end{figure}

Although the dark halo of the Galaxy extends to
very large radii ($\gg50$~kpc), mini-halos with Galactocentric
radii greater than 50~kpc are typically undetectable because
of the low cosmic-ray energy density at these radii. The
total mass of dark matter simulated is approximately
$3.3\times10^{11}\,{\rm M_\odot}$; this is only $8/11$ of the
total amount of dark matter within the simulated volume, with
the remaining $3/11$ being made up of clusters in the mass
range $10^{-1}\le M/{\rm M_\odot}<10^{2}$.

We find that a fraction $\sim1.5\times10^{-6}$ of the simulated
mini-halos have a predicted photon flux above 100~MeV of more
than $F=7\times10^{-8}\;{\rm ph\,cm^{-2}\,s^{-1}}$, thus
placing them above the approximate EGRET threshold (cf. Gehrels
et al 2001). In other words, the simulations yield $\sim280$
sources bright enough that they should appear in the EGRET
catalogue. The source counts, as a function of flux, are
shown in figure 1; the relation is clearly flatter than
usual Euclidean result ($N\propto S^{-3/2}$), and at low
flux levels is well approximated by $N\propto S^{-1}$.

The locations of the synthetic sources on the sky are
shown in figure 2, where it can be seen that although
they populate the entire sky, they are more concentrated
towards the disk of the Galaxy, and are preferentially
found toward the central regions. These features are
primarily a reflection of our adopted model of
the distribution of cosmic-rays in the Galaxy (\S4).
The median Galactic latitude of the sources is $12^\circ$.
Dividing the population into two equally-sized groups of
low and high latitude sources yields no significant
difference in median flux between the two groups.

The mass distribution of the detectable mini-halos is shown 
in figure 3. At the high mass end the sources may be
very luminous and can thus be detected to large distances,
whereas only the closest of the low mass mini-halos
are detectable. The competition between increasing mini-halo
number density and declining luminosity, towards lower
mini-halo masses, results in a characteristic (median) mass
of $5\times10^6\,{\rm M_\odot}$ for the detectable sources.
In the vicinity of the Sun, a mini-halo of this mass is
expected to have a luminosity of approximately
$8\times10^{34}\,{\rm erg\,s^{-1}}$, in photons of energy
above 100~MeV. This result is in accord with the analysis
of the UID EGRET source population by
Kanbach et al (1996), who concluded that their typical
luminosity is $\sim10^{35}\,{\rm erg\,s^{-1}}$. The median
line-of-sight distance for the mini-halos with flux greater
than the EGRET detection limit is 3.9~kpc.

Bearing in mind that both Active Galactic Nuclei (AGN) and
pulsars are effectively point-like gamma-ray sources,
the non-trivial angular size predicted for the gamma-ray
emission from dark mini-halos is an important feature
of the model. We characterise the size of the sources
by $\theta_t$, the tidal radius of the mini-halo;
this is the radius which encloses all of the flux from
the model source. We compute $\theta_t$ using the same
procedure as WOM03. For a singular isothermal sphere, the
surface brightness is $I\propto1/\theta$ at all radii,
and we expect roughly half of the flux to be contained
within $\theta_t/2$. The median value of the tidal radius
is $\langle\theta_t\rangle=2.2^\circ$, for sources
detectable by EGRET. The full distribution of $\theta_t$
is shown in figure 4.

Intrinsic source confusion, i.e. the regime where sources
are so numerous that they overlap on the sky, becomes
a minor problem in the region around the Galactic Centre.
Within $10^\circ$ latitude/longitude of the Galactic
Centre (an area of roughly $400\;{\rm deg^2}$), we find
32 sources, covering about $100\;{\rm deg^2}$ (the sources
are smaller than average). This figure is an overestimate
in the sense that our model predicts too many sources --- see
\S6. With more sensitive instrumentation, having a lower flux
limit, intrinsic source confusion becomes a major issue, as
the mini-halo population covers the entire sky (WOM03).

\section{Comparison with observations}
It is clear that the model we have presented predicts
too many EGRET sources. The entire EGRET source list
contains only about as many sources as our simulation
predicts, and many ($\sim40$\%) of the catalogued sources
are considered to be identified. Furthermore, we must
expect that there will be many examples of known types
of gamma-ray sources, such as pulsars, amongst the UID
EGRET sources, implying that our model over-predicts
the source count by a factor $\ga2$. However, it must be
acknowledged that the model we have presented involves
substantial extrapolation into poorly charted territory
and is therefore illustrative, not definitive.
There are, for example, considerable uncertainties in
the cosmic-ray distribution model, in the shape of the
Galaxy's dark halo, and in the mass spectrum of the
clustering. In addition, we can immediately point to
one aspect of the dark matter model which leads to an
over-prediction of the source counts: we have assumed
that {\it all\/} of the dark matter in the Galactic halo
is bound into mini-halos (\S2; WOM03), but this cannot be
the case. Even if all the dark matter were initially in
mini-halos, the process of tidal stripping will gradually
unbind material, leading to a significant fraction in
the form of dark matter ``streams''. Considering these
uncertainties, we recognise that the over-prediction of
source counts is not a fundamental deficiency of the model.

Other basic features of the UID source population
can be used to gauge the success of the model: the
distribution on the sky; spectra; variability; angular
structure; and observations made in other wave-bands.
We now address each of these aspects in turn.

\subsection{Sky distribution}
The model correctly predicts a preponderance of sources
at low- to mid-Galactic latitudes, with about the right
median latitude ($12^\circ$) for the population. However,
it does not exhibit the two component (thin-disk plus
thick-disk) source population which the data suggest
(Gehrels et al 2001). 

Figure 2 shows a relatively strong concentration
towards the Galactic Centre region, compared to the
data. However, in this region the predicted source
density is sufficiently high that source confusion
is expected to be a major problem with EGRET observations
of a such a population. The predicted confusion is
predominantly instrumental, because the sources are
typically unresolved (see \S6.4). We find approximately
100 sources within $\pm30^\circ$ of latitude and
longitude of the Galactic Centre -- an area of roughly
$3400\;{\rm deg^2}$. By comparison the point-spread-function
of EGRET has a Full-Width Half-Maximum (FWHM) of
approximately $4^\circ$ (see \S6.4), implying that
this entire region appears, to EGRET, to be
``covered'' with overlapping sources.

\subsection{Spectra}
The spectrum predicted by the present model is given
in Ohishi, Mori and Walker (2003): it exhibits a
peak power -- i.e. a peak in $E^2{\rm d}N/{\rm d}E$
-- at several-hundred~MeV. At this point the spectrum rolls
over from ${\rm d}N/{\rm d}E\propto E^{-1}$, at low
energies, approaching $E^{-2.75}$ at $E\gg1$~GeV.

In comparing this prediction with the data (e.g.
Merck et al 1996), there are two important points
to bear in mind. First, the model gamma-ray emission
spectrum is unique only by default: we don't know
the cosmic-ray spectrum elsewhere in the Galaxy.
The simple fact that various gamma-ray spectra
are observed should therefore not be used as an argument
against the present model. We note that the diffuse
Galactic plane emission at $E\ga1$~GeV (Hunter et al 1997)
is difficult to understand if the Galactic cosmic-ray
spectrum is everywhere the same as in the solar
neighbourhood, suggesting that the typical cosmic-ray
spectrum may be harder than measured locally (Mori 1997).

Secondly, the sources
are predicted to be extended, with low-intensity wings
on the profile extending out to $\sim2^\circ$, typically.
Although the spectrum should be uniform across the source,
the fact that the point-spread function of the detector
changes with energy, coupled with the low-level ``wings''
on the source profile, could lead to spurious estimates
of spectral shapes. In particular, some fraction of the
source flux will be absorbed into the estimate of
the background intensity, and this fraction will vary
with photon energy. These problems, coupled with the
low signal-to-noise ratio of many of the UID sources,
make it difficult to assess the success of the model
spectral predictions.

\subsection{Variability}
The model we have presented involves emission which is
intrinsically steady on observationally accessible
time-scales; it is therefore not relevant to any sources
which are known to vary significantly. Most of the UID
EGRET sources are not bright enough to permit strong
constraints on their variability. There is no consensus
in the literature regarding the variability of the UID
sources: McLaughlin et al (1996) find that only a small
fraction, roughly one in six, of the UID EGRET sources
are significantly variable (see also Wallace et al 2000);
by contrast Torres et al (2001) suggest that the fraction
may be as large as one third, in the case of low-latitude
UID sources (see also Torres, Pessah and Romero, 2001, and
Tompkins 1999). We note that large-amplitude variations do
not sit easily with the multiple/extended source designation
carried by $\sim$50\% of the UID EGRET sources (Hartman et
al 1999).

\subsection{Angular Structure}
Although the estimated angular sizes of the detectable
mini-halos are large (of order degrees), the resolving
power of EGRET is quite modest, with a point-spread-function
of $5^\circ$ Full-Width-Half-Maximum (FWHM) at 100~MeV
(Thompson et al 1993). (The in-flight calibration data
are consistent with the pre-launch calibration in respect
of the point-spread-function --- Esposito et al 1999.)
The EGRET angular resolution improves
with increasing energy, scaling roughly as $E^{-0.534}$.
However, if ${\rm d}N/{\rm d}E$ is close
to $E^{-2}$, half of the photons contributing to a source
detection are within a factor $\sim2$ of the low-energy
threshold. We therefore adopt a FWHM of $4^\circ$ as the
relevant instrumental width; sources would thus
need to be at least $8^\circ$ across in order to be
fully resolved. For an isothermal density distribution
within each mini-halo (\S5), half of the total flux
is contained within a diameter approximately equal to
$\theta_t$. Referring to figure 4 we then find that only
a tiny fraction of the synthetic population could be fully
resolved by EGRET.

If the instrumental FWHM is comparable to the source size
then the source structure will not be resolved, but the
data can nevertheless indicate that the source is extended,
by virtue of the observed intensity profile being broader
than the point spread function. Most of the synthetic sources
fall into this category. It is notable that half of the UID
EGRET sources are recorded as extended/multiple by Hartman
et al (1999). (The two possibilities cannot be differentiated
if the source structure is not fully resolved.)

\subsection{Counterparts at other wavelengths}
The basic criterion for an EGRET source to be
classified as ``UID'' is that it should not have
an obvious counterpart at other wavelengths. The
population we have modelled clearly meets this
requirement because the emission comes from
``dark'' matter.  The detectability of cold, dense
gas has been discussed by a number of authors:
Pfenniger, Combes and Martinet (1993);
Gerhard and Silk (1996); Combes and Pfenniger (1997);
Walker and Wardle (1999); and Sciama (2000). Perhaps
the simplest and most robust of
expectations is that there will be thermal emission
from the clouds, implying bright microwave sources
coincident with the gamma-ray sources. An
important point to note is that gamma-ray
and microwave luminosities are both proportional to
mini-halo mass and cosmic-ray density; consequently
the microwave flux can be estimated simply
by scaling the observed EGRET flux (Wardle and
Walker 1999; Sciama 2000).

To determine the bolometric microwave flux we
need only take the ratio of the thermal emissivity
due to cosmic-rays
($\Gamma_0\simeq10^{-4}\,{\rm erg\,s^{-1}g^{-1}}$; WOM03),
to the gamma-ray emissivity
($J=J_p+J_e\simeq5.6\times10^{-2}\;{\rm ph\,s^{-1}g^{-1}}$
above 100~MeV; \S3), yielding a microwave flux of
$S\simeq1.8\times10^{-10}F_7\;{\rm erg\,cm^{-2}s^{-1}}$.
Here the gamma-ray flux above 100~MeV is
$10^{-7}F_7\;{\rm ph\,cm^{-2}s^{-1}}$. The atmospheric
temperature is estimated to be in the range
$4.2-4.9$~K, if the individual clouds have mass
$M\sim10^{-4}-10^{-5}\,{\rm M_\odot}$ (WOM03).
The principal remaining uncertainty then lies with the
nature of the emitted spectrum: we have previously
suggested a blackbody spectrum (WOM03), whereas
Lawrence (2001) concludes that this is inconsistent
with the observed spectrum of the Blank Field SCUBA
sources (which objects he interprets in terms
of cold, planetary-mass gas clouds). More generally we can
take  $S_\nu\propto\nu^\alpha B_\nu$, with $\alpha=0$ for
blackbody emission, and $\alpha=2$ for emission from
small particles whose absorption increases in proportion
to $\nu$. Lawrence (2001) concludes that the latter
spectrum is consistent with the data on the Blank Field
SCUBA sources, providing $4.7<T({\rm K})<6.4$.
Here we restrict attention to two cases
which adequately represent the plausible range of
thermal emission spectra for the dense gas: a $T=4.2$~K
blackbody, and a $T=4.9$~K ``dusty'' ($\alpha=2$) spectrum.

Our adopted spectra differ greatly in the expected
flux at radio frequencies, where we are far below
the peak thermal emission --- see figure 5. At
low frequencies the two spectral models can be
approximated by $S_\nu\simeq4.2\,\nu^2\,F_7$~mJy
(blackbody) and
$S_\nu\simeq1.4\times10^{-5}\,\nu^4\,F_7$~mJy (dust),
with $\nu$ in GHz. However,
even in the blackbody case, where the radio emission
is relatively strong, the predicted flux would
be difficult to detect because it is so extended.
Taking the bright UID EGRET source 3EG J1835+5918
as an example (Hartman et al 1999; Mirabel et al 2000;
Reimer et al 2001), with $F_7\simeq7$, we find a predicted
flux at 1.4~GHz of roughly 60~mJy in the blackbody
case. This estimate is well above the point source detection
limit of the observations, reported by Mirabel et al (2000),
of 2.5~mJy. However, a mini-halo is certainly not point-like.
In fact it would fill the primary beam of the radio
telescope, and would be resolved-out on all but the
shortest interferometeric baselines. The very
extended sources predicted by the present model are
expected to be difficult to detect using radio
interferometers configured for high resolution imaging.

\section{Discussion}
Although the predicted microwave counterparts to UID EGRET
sources are not expected to have been detected in the
counterpart searches to date, they could be revealed
in the near future by some of the various experiments
designed to study anisotropies in the Cosmic Microwave
Background (CMB). In particular, the MAP 
satellite\footnote{{\tt http://map.gsfc.nasa.gov}}, has now
completed a sensitive all-sky survey, at several frequencies,
and the data are about to be released. Should MAP have
detected the predicted microwave counterparts? The
answer to this question depends on the spectrum of
the emission, because all of the MAP frequencies
are well below the thermal peak. Even considering the
highest frequency channel (90~GHz), the two spectral
models differ by a large factor in their predicted flux.
If the emission is blackbody, then $S_\nu\simeq20\,F_7$~Jy,
but only $S_\nu\simeq0.6\,F_7$~Jy for the ``dust'' model
(see figure 5). The point-source (i.e. single pixel)
detection limit for MAP at this frequency should be
approximately 1.7~Jy (5$\sigma$, and we have taken
the CMB contribution to be roughly equal to the thermal
noise of the instrument). On this basis we do not
expect the typical UID sources to be detected by
MAP if they have a ``dusty'' spectrum.

Even if the mini-halos have a blackbody spectrum they
might not be detected by MAP,
because its $0.3^\circ$ pixels are substantially smaller
than the predicted mini-halo sizes. For an isothermal
mini-halo density profile, the enclosed flux varies
roughly in proportion to radius. With a predicted
median source radius of $2.2^\circ$, it is evident
that the largest single-pixel flux expected from a typical
mini-halo is not much above the MAP detection limit.
Computing the peak single-pixel flux for
each source in our simulation,
we expect that only 40\% of the predicted population
ought to be detectable by MAP even if the spectra
are blackbody. However, we note that sources which
are individually undetected may still be useful for
constraining the typical microwave/gamma-ray flux
ratio of the UID EGRET population.

Other satellite and balloon-borne CMB experiments
will provide robust constraints on the model we have
presented, by virtue of observing close to the
predicted thermal peak of the mini-halos. In particular,
the High Frequency Instrument on the Planck
satellite\footnote{{\tt http://sci.esa.int/home/planck}} and
its prototype, the balloon-borne Archeops (Beno\^\i t et al 2002),
will map the sky at 353 and 545~GHz, sandwiching the point
-- around 400~GHz (see figure 5) -- where our model spectra
cross. At these frequencies the predicted flux of a typical
UID EGRET source ($F_7=1$) is of order 30~Jy, for either
spectral model. By contrast,
the limiting flux ($5\sigma$) for Planck, at 353~GHz,
will be 100~mJy in a single (5~arcmin) pixel, and
roughly 5~Jy for a source of $2.2^\circ$ radius.
(Oberving at these high frequencies has the additional
advantage of less confusion from the degree-scale CMB
anisotropies.)
Planck should therefore detect microwave counterparts
to any of the UID EGRET sources in which the gamma-rays
arise from cosmic-ray interactions with dense gas.
We note that a substantial fraction of the sky has already
been mapped by the Archeops experiment (Beno\^\i t et al 2002),
and UID EGRET source counterparts might be present in the
existing data. Counterparts are best searched for,
initially, well away from the Galactic plane as the latter
region is likely to be confused at high frequencies.
Scaling from the Planck sensitivity estimates, using
the estimated duration and coverage -- 2~year mission
with full sky coverage, for Planck, versus 24~hour
Archeops dataset covering 25\% of the sky -- we estimate
that the flux limit of Archeops, for the extended microwave
sources we predict, should be approximately 70~Jy. 
Hence UID EGRET sources brighter than 
$3\times10^{-7}\,{\rm ph\,cm^{-2}s^{-1}}$ ($>\,100$~MeV)
may have counterpart microwave sources found in
the Archeops data.

We note that the balloon-borne MAXIMA and BOOMERANG
experiments also covered frequencies close to the
thermal peak of cold gas emission, but these experiments
covered only $\sim$1\% of the sky (Hanany et al 2000;
Coble et al 2003), and these data are therefore
of limited utility in the present context. Ground-based
studies are similarly limited to small patches of the
sky, when observing at 400~GHz, but would nevertheless
be helpful if the UID EGRET source population is
specifically targetted for observations.

If microwave counterparts are discovered, then we will
be able to study the density profiles of dark matter
mini-halos directly.
Further to the properties noted in \S2 for the
individual source profiles, we can make some generic
predictions for their structure: (i) the intensity should
rise to a high central peak, but there should be a
core (Walker 1999) in the surface-brightness profile;
(ii) the limb of the source should exhibit a sharp
cut-off due to tidal truncation; and (iii) there may be
tidal streams extending along the mini-halo's orbit.

In addition to tidal streams associated with identifiable
mini-halos, it is expected that some mini-halos have
been completely disrupted by the tidal fields they have
experienced. In these cases the tidal streams are not
associated with a bound cluster, and thus represent a
distinct category of microwave/gamma-ray source predicted
by the model. Shells with sharply-defined edges could
also appear in the microwave/gamma-ray maps, as these
are a common feature of tidal debris (Hernquist and Quinn
1988; Hernquist and Spergel 1992).

The Gamma-ray Large Area Space
Telescope\footnote{{\tt http://glast.gsfc.nasa.gov}}
(GLAST) will provide a significant advance in our
understanding of the UID sources, because GLAST will
be much more sensitive than EGRET and will have better
angular resolution. These improved capabilities will
permit powerful tests of the model we have presented.
However, the data from MAP and the balloon-borne CMB
experiments will be available on a much shorter
time-scale than those from GLAST, and it should be
possible to make considerable progress with the microwave
data alone. In particular, our prediction of bright,
extended, thermal microwave counterparts is unique
amongst existing models of the UID EGRET population.

Finally we note that some of the UID EGRET sources
are sufficiently bright that they may be detectable
by ground-based TeV telescopes (Aharonian et al 1997),
even if their spectra are as steep as $E^{-2.75}$ between
the GeV and TeV bands. Studying UID sources at TeV energies
may permit their nature to be discerned. For example,
the extended nature of the sources predicted by the
present model is unusual for gamma-ray sources, and the
slightly extended (6~arcminute) TeV source, in the vicinity
of 3EG~J2033+4118, reported by Aharonian et al (2002)
is of interest in this context.

\section{Conclusions}
If our Galaxy contains a significant component of dark matter
in the form of cold, dense gas clouds, clustered into large
aggregates, some of those clusters should have been detected
by EGRET. Using a CDM-like mass spectrum for the clustering,
we have shown that the predicted gamma-ray source population
has properties which are broadly similar to those of a large
fraction of the UID EGRET sources. In particular: the Galactic
latitude distribution and the source size distribution
anticipated in the model find support in the EGRET data.
Furthermore the intrinsically ``dark'' nature of the predicted
sources naturally explains most of the difficulty in finding
counterparts; however, the large angular size of the clusters
also plays a role, because counterpart searches to date have
had poor sensitivity to $\sim\,$degree-sized sources. The
total number of UID sources predicted by the model is too
large, particularly bearing in mind that many of the observed
UID sources are likely to be examples of known types of
gamma-ray emitters. Data returned by the various CMB
experiments will provide a powerful test of the interpretation
we have presented: counterpart thermal microwaves should be
detected from the cold gas, and the source structure should
be resolved.

\acknowledgements
MAW thanks Gordon Garmire and Eric Feigelson for emphasising
the importance of the gamma-ray data, and for helpful
discussions on the relevant physics. We thank Ben Moore for
advice on modelling the mini-halo population, and Simon
Johnston for helpful comments on the manuscript. The
Referee's comments also improved the paper.


\begin{references}

\reference{} Aharonian~F.A., Hofmann~W., Konopelko~A.K., V\"olk~H.J. 1997 APh 6, 369
\reference{} Aharonian~F.A. et al 2002 A\&A 393, L37

\reference{} Bailes~M., Kniffen~D.A. 1992 ApJ 391, 659
\reference{} Benaglia~P., Romero~G.E., Stevens~I.R, Torres~D.F. 2001 A\&A 366, 605
\reference{} Beno\^\i t~A. et al 2002 APh 17, 101
\reference{} Bergstr\"om~L., Edsj\"o~J., Gunnarsson~C. 2001 PhRvD 63(8), 3515
\reference{} Bloemen~H. 1989 ARAA 27, 469
\reference{} Blumenthal~G., Faber~S.M., Primack~J.R., Rees~M.J. 1984 Nature 311, 527

\reference{} Calc\'aneo-Rold\'an~C., Moore~B. 2000 PhRvD 62(12), 3005
\reference{} Combes~F., Pfenniger~D. 1997 A\&A 327, 453
\reference{} Coble~K. et al 2003 ApJS (Submitted) (astro-ph/0301599)

\reference{} Davis~M., Efstathiou~G., Frenk~C.S., White~S.D.M. 1985 ApJ 292, 371

\reference{} De~Paolis~F., Ingrosso~G., Jetzer~Ph., Roncadelli~M. 1995
             A\&A, 295, 567
\reference{} De~Paolis~F., Ingrosso~G., Jetzer~Ph., Roncadelli~M. 1999
             ApJL, 510, L103
\reference{} Dixon~D.D., Hartmann~D.H., Kolaczyk~E.D., Samimi~J., Diehl~R.,
             Kanbach~G., Mayer-Hasselwander~H., Strong~A.W. 1998 NewAst 3, 539
\reference{} Esposito~J.A. et al 1999 ApJS 123, 303

\reference{} Gehrels~N., Macomb~D.J., Bertsch~D.L.,
             Thompson~D.J., Hartman~R.C. 2001 Nature, 404, 363
\reference{} Gerhard~O., Silk~J. 1996, ApJ, 472, 34
\reference{} Ghigna~S., Moore~B., Governato~F., Lake~G., Quinn~T., Stadel~J. 1998 MNRAS 300, 146
\reference{} Gregory~P.C., Vavasour~J.D., Scott~W.K., Condon~J.J. 1994 ApJS 90, 173
\reference{} Grenier~I.A., Perrot~C.A. 2001 AIP Conf. Proc. 587, 649

\reference{} Hanany~S. et al 2000 ApJL 545, L5
\reference{} Hartman~R.C. et al 1999 ApJS 123, 79
\reference{} Henriksen~R.N., Widrow~L.M. 1995, ApJ, 441, 70
\reference{} Hernquist~L., Quinn~P.J. 1988 ApJ 331, 682
\reference{} Hernquist~L., Spergel~D.N. 1992 ApJL 399, L117
\reference{} Hogan~C.J. 1993 ApJL 415, L63
\reference{} Hunter~S.D. et al 1997 ApJ 481, 205

\reference{} Kaaret~P., Cottam~J. 1996 ApJL 462, L35
\reference{} Kalberla~P.M.W., Shchekinov~Yu.A., Dettmar~R.-J. 1999 A\&A 350, L9
\reference{} Kanbach~G. et al 1996 A\&AS 120, 461
\reference{} Kauffmann~G., White~S.D.M., Guideroni~B. 1993 MNRAS 264, 201
\reference{} Klypin~A., Kravtsov~A.V., Valenzuela~O., Prada~F. 1999 ApJ 522, 82

\reference{} Lawrence~A. 2001 MNRAS 323, 147

\reference{} McLaughlin~M.A., Mattox~J.R., Cordes~J.M.,
             Thompson~D.J. 1996 ApJ 473, 763
\reference{} Merck~M. et al 1996 A\&AS 120, 465
\reference{} Mirabel~N., Halpern~J.P., Eracleous~M., Becker~R.H. 2000 ApJ 541, 180
\reference{} Montmerle~T. 1979 ApJ 231, 95
\reference{} Moore~B., Ghigna~S., Governato~F., Lake~G., Quinn~T.,
             Stadel~J., Tozzi~P. 1999 ApJL 524, L19
\reference{} Mori~M. 1997 ApJ 478, 225

\reference{} Ohishi~M., Mori~M, Walker~M.A. 2003 (In preparation)

\reference{} Peebles~P.J.E. 1993 ``Principles of Physical Cosmology''
             (PUP, Princeton)
\reference{} Pfenniger~D., Combes~F., Martinet~L. 1994, A\&A, 285, 79
\reference{} Porter~T.A., Protheroe~R.J. 1997 Nucl. Part. Phys. 23, 1765

\reference{} Rafikov~R.R., Draine~B.T. 2001 ApJ 547, 207
\reference{} Reimer~O., Brazier~K.T.S., Carrami\~nana~A., Kanbach~G.,
             Nolan~P.L., Thompson~D.J. 2001 MNRAS 324, 772
\reference{} Roberts~M.S.E., Romani~R.W., Johnston~S. 2001 ApJL 561, L187

\reference{} Sciama~D.W. 2000 MNRAS 312, 33
\reference{} Skibo~J.G., Ramaty~R. 1993 A\&AS 97, 145
\reference{} Strong~A.W., Moskalenko~I.V., Reimer~O. 2000 ApJ 537, 763
\reference{} Sturner~S.J., Dermer~C.D., Mattox~J.R. 1996 A\&AS 120, 445

\reference{} Thompson~D.J. et al 1993 ApJS 86, 629
\reference{} Tompkins~W.F. 1999 PhD Thesis (Stanford University)
             (astro-ph/0202141)
\reference{} Torres~D.F., Pessah~M.E., Romero~G.E. 2001 Ast. Nachr. 322, 223
\reference{} Torres~D.F., Romero~G.E., Combi~J.A., Benaglia~P.,
             Andernach~H., Punsly~B. 2001 A\&A 370, 468
\reference{} Torres~D.F., Romero~G.E., Dame~T.M., Combi~J.A.,
             Butt~Y.M. 2003 PhRep, Submitted (astro-ph/0209565)
\reference{} Turner~M.S., Tyson~J.A. 1999 RModPhys 71, S145

\reference{} Walker~M.A. 1999 MNRAS 308, 551
\reference{} Walker~M.A., Lewis~G.F. 2003 ApJ, Submitted
             (astro-ph/0212345)
\reference{} Walker~M.A., Ohishi~M., Mori~M. 2003 ApJ, Submitted
             (astro-ph/0210483) (WOM03)
\reference{} Walker~M., Wardle~M. 1998 ApJ 498, L125
\reference{} Walker~M., Wardle~M. 1999 PASA 16, 262
\reference{} Wallace~P.M., Griffis~N.J., Bertsch~D.L., Hartman~R.C.,
             Thompson~D.J., Kniffen~D.A., Bloom~S.D. 2000 ApJ 540, 184
\reference{} Wardle~M., Walker~M. 1999 ApJL 527, L109
\reference{} Webber~W., Lee~M., Gupta~M. 1992 ApJ 390, 96 (WLG92)

\reference{} Yadigaroglu~I.-A., Romani~R.W. 1997 ApJ 476, 347

\end{references}
\end{document}